\documentstyle[12pt]{article}

\begin{document}
\begin{center}
{\Large \bf On the evolution law of the universe}
\bigskip

{\large D.L.~Khokhlov}
\smallskip

{\it Sumy State University, R.-Korsakov St. 2, \\
Sumy 244007, Ukraine\\
E-mail: others@monolog.sumy.ua}
\end{center}

\begin{abstract}
The model of the homogenous and isotropic universe
is considered in which the coordinate system of reference
is not defined by the matter but is a priori specified.
The scale factor of the universe changes following
the linear law. The scale of mass changes proportional to
the scale factor of the universe. The model under consideration
avoids the flatness and horizon problems.
The predictions of the model is fitted to the observational
constraints: Hubble parameter, age of the universe and CMB data.
\end{abstract}

\section{Introduction}

As known~\cite{Dol,Lin}, the Friedmann
model of the universe has fundamental difficulties such as the
flatness and horizon problems due to the slow growth
of the scale factor of the universe.
In the Friedmann universe, the coordinate system of reference is
associated with the matter of the universe.
The evolution of the scale factor of the universe is given by
\begin{equation}
a \sim t^{1/2}, \qquad
a \sim t^{2/3}
\label{eq:a1}
\end{equation}
where the first equation corresponds to the matter as
a relativistic gas, and the second, to the dust-like matter.
Growth of the scale factor of the universe governed by
the power law with the exponent less than unity
is slower than growth of the horizon of the universe
$h \sim t$
that causes the flatness and horizon problems.

To resolve these problems an inflationary episode in the early
universe is introduced~\cite{Dol,Lin}. However there is another
way of resolving the problems. This is based on the premise
that the coordinate system of reference is not
defined by the matter but is a~priori specified.

\section{Theory}

Let us consider the model of the homogenous and isotropic
universe.
Let us assume that the coordinate system of reference is not
defined by the matter but is a~priori specified.
Let the coordinate system of reference be the Euclidean space
with the spatial metric $dl$ and absolute time $t$
\begin{equation}
dl^2=a(t)^2(dx^2+dy^2+dz^2), \quad t.
\label{eq:met}
\end{equation}
That is the coordinate system of reference is the
space and time of the Newton mechanics, with the scale factor of
the universe is a function of time.
Since the metric (\ref{eq:met}) is not defined by the matter,
we can a~priori specify the evolution law of the scale factor
of the universe. Let us take the linear law when
the scale factor of the universe grows with the velocity of light
\begin{equation}
a=ct.\label{eq:g1}
\end{equation}

In the Friedmann universe, the law~(\ref{eq:g1})
corresponds to the Milne model~\cite{Zeld}
which is derived from the condition that the density of the matter
tends to zero $\rho\rightarrow 0$. Here Eq.~(\ref{eq:g1})
describes the universe in which the evolution of the scale factor
do not depend on the presence of the matter. Hence the
density of the matter is not equal to zero $\rho\not= 0$.
The total mass of the universe relative to the background space
includes the mass of the matter and the energy of its gravity.
Let us adopt that the total mass of the universe is equal to zero,
that is
the mass of the matter is equal to the energy of its gravity
\begin{equation}
c^2={Gm\over{a}}.\label{eq:o}
\end{equation}
Allowing for Eq.~(\ref{eq:g1}),
from Eq.~(\ref{eq:o}) it follows that
the mass of the matter changes with time as
\begin{equation}
m={c^2a\over{G}}={c^3t\over{G}},\label{eq:p}
\end{equation}
and the density of the matter, as
\begin{equation}
\rho={{3c^2}\over{4\pi G a^2}}={3\over{4\pi G t^2}}.\label{eq:q}
\end{equation}
Hence the model under consideration yields the change of the scale
of mass proportional to the scale factor of the universe.
At the Planck time $t_{Pl}=(\hbar G/c^5)^{1/2}$,
the mass of the matter is equal to the Planck mass
$m_{Pl}=(\hbar c/G)^{1/2}$. At present, the mass of the matter is
of order of the modern value $m_0={c^2a_0/{G}}$.

Let us consider the flatness and horizon problems within the
framework of the model under consideration.
Remind~\cite{Dol,Lin} that the horizon problem
in the Friedmann universe is caused by that the universe
observable at present
consisted of the causally unconnected regions in the past
that is inconsistent with the high isotropy of the background
radiation.
In the universe under consideration,
the size of the universe
(the scale factor of the universe)
coincides with the size of the horizon
during all the evolution of the universe.
Hence the presented model avoids the horizon problem.

Remind \cite{Dol,Lin} that
the essence of the flatness problem
in the Friedmann universe
is connected with impossibility to gain the
modern density of the matter at present starting from
the Planck density of the matter at the Planck time.
In the presented theory,
the density of the matter of the universe
changes from the Plankian value at the Planck time
to the modern value at the modern time.
Hence the flatness problem is absent in the presented theory.

\section{Predictions}

In view of Eq.~(\ref{eq:g1}),
we write down some relations describing the universe.
The relation between the Hubble parameter and the age
of the universe is given by
\begin{equation}
H=\frac{1}{a}\frac{da}{dt}=\frac{1}{t}.\label{eq:j}
\end{equation}
The scale factor of the universe at time $z$ is given by
\begin{equation}
a(z)=\frac{a_{0}}{z+1}.\label{eq:sf}
\end{equation}
The age of the universe at time $z$ is given by
\begin{equation}
t(z)=\frac{1}{H_{0}(z+1)}.\label{eq:au}
\end{equation}
The angular diameter distance at time $z$ is given by
\begin{equation}
d(z)=\frac{c}{H_{0}}\frac{z}{z+1}.\label{eq:ad}
\end{equation}

Let us fit the predictions of the model to the observational
constraints: Hubble parameter, age of the universe and CMB data.
The Hubble parameter deduced from the HST observations of
Cepheids are converging to a value~\cite{Bur}
$H_0=63\pm 10{\ \rm \ km/c/Mpc}$.

The age of the universe is estimated by
the observed age of the oldest globular clusters
$t_{GC}=14\pm 2 {\ \rm Gyr}$~\cite{Cow}.
According to Eq.~(\ref{eq:au}),
for $H_0=63{\ \rm \ km/c/Mpc}$,
the age of the universe is
$t_0=15.5{\ \rm Gyr}$.

Consider classical angular diameter distance test
with the standard ruler provided by the
cosmic microwave background (CMB) temperature anisotropy
power spectrum~\cite{Hu}.
Any feature projects as an anisotropy onto
an angular scale associated with multipole
\begin{equation}
\ell=kr,\label{eq:m}
\end{equation}
where $k$ is the size of the feature in $k$ space,
$r$ is the distance to the feature.
Last scattering in the epoch of recombination
$z_{ls}=1400$ projects the observed multipole $\ell=260$~\cite{Line}.
Usually the standard ruler is taken as the sound horizon
at last scattering,
and the distance to the ruler is an angular diameter distance
to last scattering.
Let us assume that the standard ruler is
the size of the universe at last scattering
\begin{equation}
k_{ls}=\frac{H_{0}(z_{ls}+1)}{c}.\label{eq:rls}
\end{equation}
The distance to the ruler is the sound horizon
at last scattering measured at present
\begin{equation}
r_{ls}=\frac{d_{ls}}{\pi\sqrt{3}}.\label{eq:dls}
\end{equation}
Substituting Eqs.~(\ref{eq:ad}), (\ref{eq:rls}), (\ref{eq:dls})
into Eq.~(\ref{eq:m})
we obtain
\begin{equation}
\ell_{ls}=\frac{z_{ls}}{\pi\sqrt{3}}=257.\label{eq:mls}
\end{equation}

Let us estimate the rate of growth of density fluctuations
within the framework of the Newton gravity.
Density fluctuations change with time as~\cite{Zeld}
\begin{equation}
\delta=exp(\int\sqrt{4\pi G\rho}dt).
\label{eq:de}
\end{equation}
Substituting Eq.~(\ref{eq:q}), we obtain
\begin{equation}
\delta=exp(\sqrt{3}{\rm ln}t)=t^{\sqrt{3}}.
\label{eq:de1}
\end{equation}
Let us determine the CMB anisotropy via the density fluctuations
at the moment of recombination $z_{ls}=1400$
\begin{equation}
\frac{\Delta T}{T}=\frac{1}{3}\delta(z_{ls}).
\label{eq:an}
\end{equation}
If $\delta\sim 1$ corresponds to the formation of galaxy clusters
in the epoch of reionization $z_{i}=5$, then according to
Eq.~(\ref{eq:de1}),
$\delta(z_{ls})=7.8\cdot 10^{-5}$. This gives
the CMB anisotropy $\Delta T/T=2.6\cdot 10^{-5}$.
COBE satellite~\cite{Ben} detected fluctuations in the CMB
on the large scale at the level of
$\Delta T/T\sim 10^{-5}$.

\end{document}